\documentclass[twocolumn]{revtex4}
\usepackage[dvips]{graphicx}
\usepackage{amssymb}
\usepackage{amssymb}
\usepackage{pifont}

\begin{document}

\title{Damping by slow relaxing rare earth impurities  in Ni$_{80}$Fe$_{20}$}

\author{G.~Woltersdorf$^1$}
\author{M.~Kiessling$^1$}
\author{G.~Meyer$^2$}
\author{J.-U.~Thiele$^2$}
\author{C.~H.~Back$^1$}

\affiliation{$^1$  Department of Physics, Universit{\"a}t
Regensburg, 93040 Regensburg, Germany} \affiliation{$^2$San Jose
Research Center, Hitachi Global Storage Technologies, U.S.A.}

\begin{abstract}
Doping Ni$_{80}$Fe$_{20}$ by heavy rare earth atoms alters the
magnetic relaxation properties of this material drastically. We
show that this effect can be well explained by the slow relaxing
impurity mechanism. This process is a consequence of the
anisotropy of the on site exchange interaction between the 4f
magnetic moments and the conduction band. As expected from this
model the magnitude of the damping effect scales with the
anisotropy of the exchange interaction and increases by an order
of magnitude at low temperatures. In addition our measurements
allow us to determine the relaxation time of the 4f electrons as a
function of temperature.
\end{abstract}

\maketitle

The dynamic response of magnetic materials is of fundamental
interest as well as essential for various applications in modern
magnetic storage technology. Often it is desirable to tailor the
damping parameter and the resonance frequency of magnetic
materials independently. While the resonance frequency can be
controlled relatively easily by e.g.~controlling the saturation
magnetization it is more difficult to change the Gilbert damping
parameter in a controlled way. Recent experiments have
demonstrated the ability to modify the damping parameter of
transition metals and transition metal alloys by introducing rare
earth (RE) impurities \cite{Bailey-IEEE01,Reidy-APL03} or 3d- and
5d transition metals \cite{Rantschler-JAP06}. Unfortunately a
convincing  microscopic understanding of the origin of RE-induced
damping in metallic alloys is still missing.

Here, we present experimental results on the magnetization
dynamics of thin Ni$_{80}$Fe$_{20}$ films doped with the
lanthanides Gd, Tb, Dy, and Ho. By varying the dopant
concentration we were able to tune the damping parameter by two
orders of magnitude. The dynamic response was measured over a wide
frequency range (0.5-35 GHz). By employing various resonance
techniques and configurations extrinsic and intrinsic relaxation
effects are separated. This procedure allows us to precisely
determine the induced damping for the various rare earth dopants
and, more importantly, to unambiguously identify the physical
origin if the effect. We find that the slow relaxation of the 4f
electron spins of the rare earth atoms is responsible for the
induced damping. As this mechanism should also lead to a very
strong temperature dependence of the relaxation we perform
temperature dependent FMR measurements to test the applicability
of the slow relaxer model.

A series of 10 and 30 nm thick RE doped Ni$_{80}$Fe$_{20}$ films
was grown by DC magnetron co-sputtering from single element
targets. A 1~nm thick Ta seed layer was first deposited onto the
glass substrates. The RE doped Ni$_{80}$Fe$_{20}$ films were
capped with a  3~nm thick Ta layer to prevent oxidation. During
deposition an Ar gas pressure of $2\times 10^{-3}$~mbar was used
and the deposition rate was about 0.1~nm/s. The film thicknesses
of all samples were measured by x-ray reflectivity, and the RE
concentration was determined by Rutherford Back Scattering; the
uncertainty of this
method is below 1~at.~\%. 
The static properties of the samples where investigated by
vibrating sample magnetometry and magneto optic Kerr effect
 measurements. In Ni$_{80}$Fe$_{20}$-RE earth intermetallic
alloys the 4f magnetic moments of the RE atoms are coupled to the
3d of the Ni$_{80}$Fe$_{20}$ moments via the intra-atomic 5d
orbitals \cite{Campbell, Coey}. The resulting effective 5d-3d
exchange coupling is antiferromagnetic (AF) leading to an
antiparallel alignment of RE 4f moments and the Ni$_{80}$Fe$_{20}$
3d  moments  \cite{Campbell, Coey}. This ferrimagnetic order leads
to a decreasing saturation magnetization with increasing RE
content. In our samples we observe a linear decrease of the
saturation magnetization as a function of doping for all RE
elements with a slope of about 40~Oe per atomic percent of RE
doping at room temperature (RT). All samples have soft magnetic
properties with small coercive fields (less than 2 Oe) and small
uniaxial anisotropy fields (less than 5 Oe). From a structural
point of view all samples discussed in this letter exhibit a
polycrystalline fcc structure typical for low RE concentrations
(below 8\%) used here \cite{Bailey-IEEE01}.
However clustering of the RE atoms does not occur even in the
amorphous phase at much higher RE concentrations \cite{Robinson}.

\begin{figure}
  \vskip -0.8cm
    \begin{center}
          \includegraphics[scale=0.70]{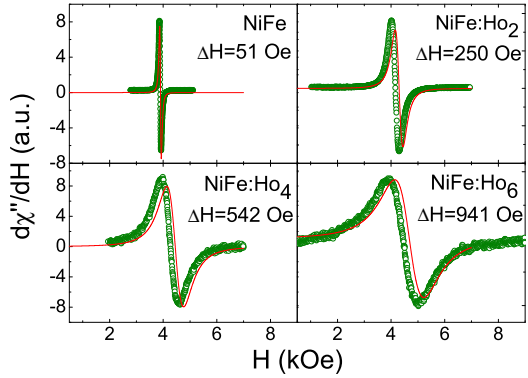}
    \end{center}
    \vskip -1cm
 \caption{FMR spectra for four different Ho doping concentrations measured at room
 temperature in the in the in-plane configuration
 using a frequency of 22 GHz. The red lines show the expected FMR lines
given by the saturation magnetization, the g-factor, and the
damping constant. In the calculation the upshift of the resonance
field with increasing doping is caused by the decreasing
magnetization and increasing damping constant. The g-factor was
determined by out of plane FMR measurements and remains nearly
unchanged up to a doping level of 6~\%. Note that for the 6~\%
sample the discrepancy between the expected and the measured line
position is about 300~Oe.} \label{dh-ho}
\end{figure}
In ferromagnetic resonance measurements the  linewidth $\Delta
H(f)$ is proportional to the microwave frequency $f$ only for
Gilbert damping. Two magnon scattering due to defects and the
superposition of local resonance lines due to large scale magnetic
inhomogeneities lead to a zero frequency linewidth $\Delta H(0)$
 \cite{Heinrich-JAP85,McMichael-PRL03}.
If  Gilbert damping dominates one has $\Delta H(0)\ll\Delta H$ and
the linewidth at a given frequency can be converted into the
damping parameter using $\alpha=\Delta H \gamma/\omega$.
\begin{figure}
 \vskip -0.8cm
    \begin{center}
\hskip -0.5cm
         \includegraphics[scale=0.74]{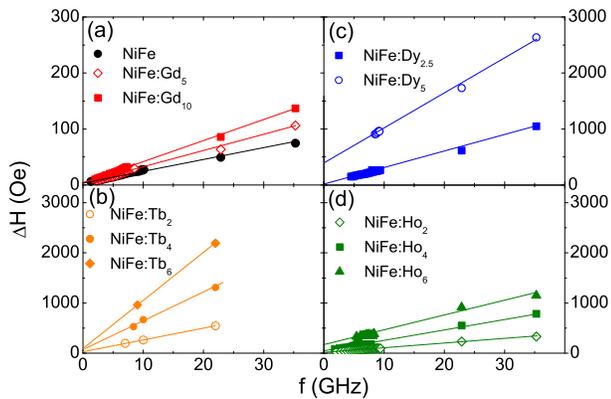}
    \end{center}
    \vskip -1cm
 \caption{Frequency dependence of  $\Delta H$ for (a) Ni$_{80}$Fe$_{20}$  and Gd
doped Ni$_{80}$Fe$_{20}$ (b) Tb doped Ni$_{80}$Fe$_{20}$ (c) Dy
doped Ni$_{80}$Fe$_{20}$  and (d) Ho doped Ni$_{80}$Fe$_{20}$
films. All measurements are carried out in the parallel
configuration at RT.} \label{dh}
\end{figure}
The series of  FMR lines shown in Fig.~\ref{dh-ho} was measured at
a frequency of 22 GHz using Ho-concentrations of $0-6$~\%. This
data illustrates the broadening of the FMR linewidth as a function
of the rare earth concentration. The pure Ni$_{80}$Fe$_{20}$ film
exhibits a linewidth of approximately 50~Oe whereas $\Delta H$
increases  by a factor of 20 to 940~Oe for a doping concentration
of 6~\%~Ho.

The observed linewidth broadening can generally  have various
origins. In order to be able to distinguish contributions from (i)
Gilbert damping, (ii) two magnon scattering and (iii) sample
inhomogeneity we perform FMR measurements over a wide frequency
range allowing us to estimate $\Delta H(0)$ and thus to verify
whether significant extrinsic contributions are present.
Out-of-plane FMR measurements further allow one to separate
magnetic inhomogeneity (local resonance) and two magnon scattering
contributions \cite{Arias-PRB99,McMichael-PRL03}, as for
 inhomogeneous samples one expects a  line broadening in
the perpendicular configuration compared to the parallel
configuration ($\Delta H_{\perp}>\Delta H_{\parallel}$)
\cite{bland-Spinger05}. For all samples discussed in this
manuscript the out of plane angular dependence of $\Delta H$ (not
shown) is consistent with Gilbert damping and we observe $\Delta
H_{\perp}\approx\Delta H_{\parallel}$.

\begin{figure}
  \vskip -0.8cm
    \begin{center}
         \includegraphics[scale=0.7]{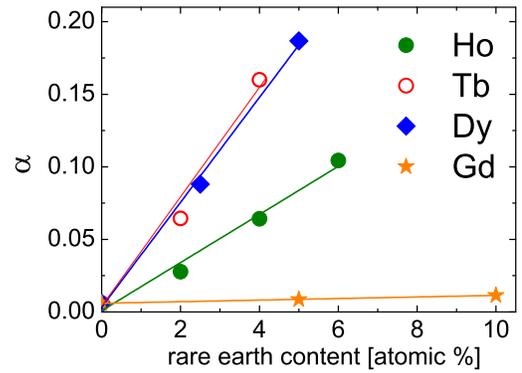}
    \end{center}
    \vskip -1cm
 \caption{ Damping parameter as a function of the RE
concentration at RT for RE=Gd, Tb, Dy, and Ho measured at RT. }
\label{alph}
\end{figure}

Fig.~2 shows the FMR linewidth as a function of frequency for
various concentrations of Gd, Tb, Dy, and Ho. The linewidth
strongly increases with increasing Tb, Dy, and Ho doping
concentration, while almost no effect is observed for Gd doping
(note the $10\times$ reduced scale for Gd doping). For all films
at doping levels of 6~\% and below the linewidth at zero frequency
$\Delta H(0)$ is very small compared to the total linewidth at 22
GHz and can be neglected. We conclude that for doping
concentrations up to 6~\% for all RE dopants the parameter
$\alpha$ can be easily determined from the slope of the frequency
dependent linewidth. The
 results are  summarized in Fig.~\ref{alph}.
For Gd doping of 5~$\%$ the damping constant of the
Ni$_{80}$Fe$_{20}$ film is not considerably influenced. On the
other hand Ho, Tb and Dy doped Ni$_{80}$Fe$_{20}$  films show a
very strong dependence of the damping parameter on the dopant
concentration. With increasing RE concentration the damping
parameter $\alpha$ increases linearly. From the slope of the
linear increase we  determine the contribution to the damping
parameter per concentration of RE dopant, i.e.~$\alpha=\alpha_{\rm
NiFe}+\Delta\alpha_{\rm RE} C_{\rm RE}$, where $C_{\rm RE}$ is the
RE atomic concentration in percent. The values for
$\Delta\alpha_{\rm RE}$ are 0.0005, 0.038, 0.036, 0.017 for Gd,
Tb, Dy, and Ho. We observe that $\Delta\alpha_{\rm RE}$ for Tb and
Dy doping are similar and lead to the largest damping, while the
value for Ho is only about half that of Tb and Dy. For Gd doping
the contribution is very small and only a consequence of the
reduced magnetization.
\begin{figure}
 \vskip-1cm
    \begin{center}
       \includegraphics[scale=0.7]{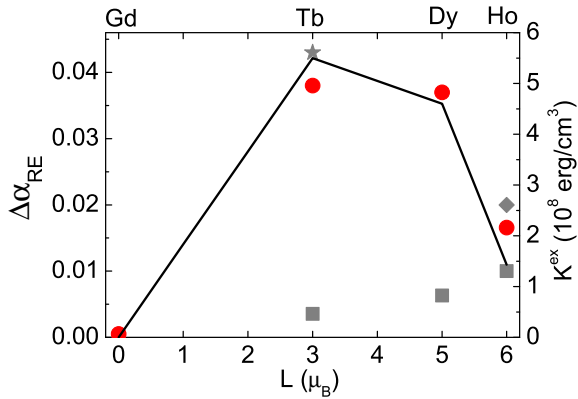}
    \end{center}
    \vskip-0.9cm
 \caption{Comparison of the RE contributed damping $\Delta \alpha_{\rm RE}$ (red bullets)
to the anisotropy contribution arising from the anisotropic
exchange interaction between the 4f moments and the conduction
electrons $K^{\rm ex}$(black line) vs.~L \cite{Irhkin88}. In
addition values of RE induced damping in Ni$_{80}$Fe$_{20}$ are
plotted from Refs. \cite{Reidy-APL03} (gray squares),
\cite{Bailey-IEEE01,Russek-JAP02} (gray star), and
\cite{MMM-seagate} (gray diamond). In Ref.~\cite{Reidy-APL03} the
damping is given as relaxation rates $\lambda$ which were
converted into the damping parameter alpha using
$\alpha=\lambda/(\gamma 4\pi M)$. } \label{irkin}
\end{figure}
This is a striking observation considering that earlier
experimental \cite{Reidy-APL03} and theoretical \cite{Rebei-PRL03}
studies suggested RE induced damping to be proportional to the
orbital moment of the dopants. Based on these predictions one
should observe the largest effect for Ho (L=6) doping, a smaller
effect for Dy (L=5) and Tb (L=3), and no effect for Gd (L=0). This
behavior is not observed in our detailed measurements. In
Fig.~\ref{irkin} the contributed damping parameter
$\Delta\alpha_{\rm RE}$ is plotted as a function of the orbital
moment of the dopants. Clearly the contributed damping is not
proportional to the orbital moment of the dopants.

In the 1960s Orbach and van Vleck \cite{vanVleck-PRL63} introduced
the 'slow relaxing' (SR) impurity model to describe the damping in
RE-doped Yttrium Iron Garnets (YIG)\cite{Seiden-PR63}. The essence
of the slow relaxer model is the following: the 4f multiplet of
the RE is split in the moderate exchange field of the 5d
electrons. This Zeeman splitting is of the order of 10 meV and the
levels are hence thermally populated at RT. In The anisotropy of
the 4f-5d exchange interaction causes a modulation of the 4f
exchange splitting when the 3d moments precess. The thermal
population of the 4f levels follows this temporal modulation, but
is delayed by the RE relaxation time $\tau_{\rm RE}$. Thermal
transitions in the 4f multiplet lead to a locally fluctuating
transverse field $h(t)$ acting on the 3d moments via the strong
3d-5d coupling if the 4f-5d exchange interaction is anisotropic.
 Using the second
fluctuation dissipation theorem one can show that the damping
constant is given by the Fourier transform of the time correlation
of $h(t)$ \cite{mori}. The time correlation function of the
fluctuation field can be approximated by $\langle h(t)h(0)\rangle
=h^2\exp{-t/\tau}$ leading to the following expression for the
Gilbert damping parameter:
\begin{eqnarray}
\label{emori} \alpha_{\rm RE}= C\times F(T) \times
\left[\frac{\tau_{\rm RE}}{1+(\omega\tau_{\rm
RE})^2}+i\frac{\omega\tau_{\rm RE}^2}{1+(\omega\tau_{\rm
RE})^2}\right]
\end{eqnarray}
where the constant $C$ is given by $C=\frac{A C_{\rm RE}}{6 M_S
k_B T}$ and the temperature dependent function  $F(T)$ accounts
for the fact that the precession induced repopulation of the 4f
levels strongly depends on the temperature \cite{Sparks-JAP67}.
$M_S$ is the saturation magnetization, $C_{\rm RE}$ is the
concentration of the RE ions, and $A$ is the anisotropy of the
5d-4f exchange interaction and given by the angular derivatives of
the 5d-4f exchange energy $\hbar\Omega$ \cite{Sparks-JAP67}. Its
magnitude can be estimated from the anisotropy contribution
arising from the anisotropic exchange interaction between the 4f
moments and the conduction electrons as observed in metallic rare
earth single crystals $K^{\rm ex}$ \cite{Irhkin88}.
 For a two level
system one has $F(T)={\rm sech}^2\frac{\hbar \Omega}{k_B T}$
\cite{Sparks-JAP67}. The population of the 4f levels in RE doped
YIG is indeed well described as a two level system due to the
large crystal field spitting. In experiments both $\Delta H_{\rm
RE}$ and $S_{\rm RE}$ have shown a strong temperature dependence
with a peak occurring at low T when $\hbar \Omega=k_BT$
\cite{Seiden-PR63,Sparks-JAP67,Clarke-PR65}. The two level
approximation, however, may not be justified for RE doped
Ni$_{80}$Fe$_{20}$. Due to the absence of a significant crystal
field in the metallic alloy 2J+1 4f levels need to be considered
allowing transitions to occur at 2J different energies. Therefore
considerable broadening of the linewidth peak at low temperature
(compared to RE doped YIG) is expected for RE doped
Ni$_{80}$Fe$_{20}$.

The real part of Eq.\ref{emori} corresponds to damping and causes
a linewidth $\Delta H_{\rm RE}=\Re(\alpha)\frac{\omega}{\gamma}$
while the imaginary part leads to a negative field shift  $S_{\rm
RE}=-\Im(\alpha)\frac{\omega}{\gamma}$. Provided that $\omega \ll
1/\tau_{\rm RE}$ the damping is independent of the frequency and
the resonance field shift is small. The negative field shift is a
consequence of the time delayed damping torque (due to thermal
repopulation of the 4f levels) leading to an effective
longitudinal field.
\begin{figure}
\vskip -1cm
    \begin{center}
         \includegraphics[scale=0.65]{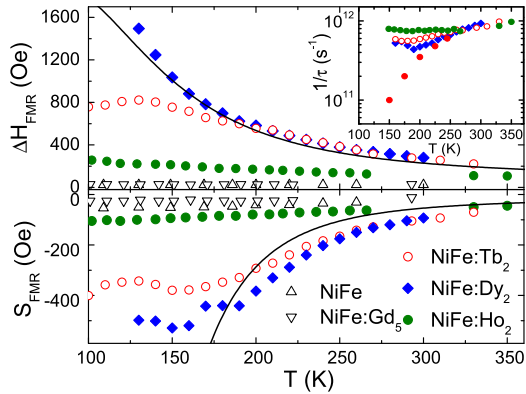}
    \end{center}
 \vskip-0.9cm
 \caption{Temperature dependence of the FMR linewidth ($\Delta H$)  and the FMR
field shift $S$ for a dopant concentration of 2.0~\%  for Tb and
Ho, 2.5~\% for Dy, and 5~\% for Gd. The measurements where carried
out at 10 GHz in the parallel configuration. The field shift $S$
corresponds to the difference between the resonance fields of the
doped and undoped samples. Note that the data for 2.5~\% Dy were
multiplied by 4/5 in order to be able to compare them to the Tb
and Ho 2.0~\% data. The solid lines represent the expected
behavior and were calculated from Eq.~\ref{emori}. For the
temperature dependence $\tau_{RE}$ we used Eq.~13 of
\cite{Clarke-PR65}. The inset shows the temperature dependence of
$\tau_{\rm RE}$ as determined from the present FMR measurements.
The red bullets represents earlier measurement of $\tau_{\rm RE}$
for Nd doped YIG \cite{Clarke-PR65}.} \label{temp}
\end{figure}
In order to predict the relative strength of this effect at a
given temperature as a function of the RE element one only needs
to compare the calculated relative magnitude of the anisotropic
exchange contribution to the magnetic anisotropy \cite{Irhkin88}
as shown in fig.\ref{irkin}. One finds: $K^{\rm ex}_{\rm Gd}=0$,
$K^{\rm ex}_{\rm Tb}=5.5\times 10^8$ erg/cm$^3$, $K^{\rm ex}_{\rm
Dy}=4.6\times 10^8$ erg/cm$^3$, and $K^{\rm ex}_{\rm
Ho}=1.43\times 10^8$ erg/cm$^3$ \cite{Irhkin88}. Neglecting the
temperature dependence in Eq.~\ref{emori} (it should be roughly
the same for the various RE elements) one immediately observes
that the RE induced relaxation should be significantly smaller for
Ho dopants compared to Tb and Dy. It is also  apparent from this
analysis why doping with Gd with its isotropic 4f-5d exchange
interaction (S-state) cannot lead to additional relaxation.

The applicability of the SR model for RE doped Ni$_{80}$Fe$_{20}$
can be further tested by verifying whether $\Delta H_{\rm RE}$ and
$S_{\rm RE}$ increase with decreasing temperature as predicted by
Eq.\ref{emori}. Fig.~\ref{temp}a shows the temperature dependence
$\Delta H_{\rm RE}$ and $S_{\rm RE}$ measured at $f$=10~GHz for
Ni$_{80}$Fe$_{20}$, Ni$_{80}$Fe$_{20}$:Gd$_{5}$,
Ni$_{80}$Fe$_{20}$:Tb$_{2}$, Ni$_{80}$Fe$_{20}$:Dy$_{2.5}$, and
Ni$_{80}$Fe$_{20}$:Ho$_{2}$. Indeed for Tb, Dy, and Ho, the data
can be well described by Eq.~\ref{emori}.
 The expected negative field shift of the
resonance field is clearly observed, see  Fig.~\ref{temp}. The
temperature dependence of  $\Delta H_{\rm RE}$ and $S_{\rm RE}$ is
expected to be similar since it is primarily caused by F(T) and
$\tau_{\rm RE}(T)$. Using Eq.~\ref{emori} the RE relaxation time
$\tau_{\rm RE}$ can be estimated from the ratio $2|S_{\rm
RE}|/\Delta H_{\rm RE}=\omega\tau_{\rm RE}$. From the data shown
in Fig.~\ref{temp} for $S_{\rm RE}$ and $\Delta H_{\rm RE}$ one
can estimate at room temperature $\tau^{300 \ \rm K }_{\rm RE}\sim
1$~ps and at low temperature $\tau^{120 \ \rm K }_{\rm RE}\sim
3-10$~ps for Tb, Dy and Ho doping.  This is in excellent agreement
with earlier independent measurements \cite{Clarke-PR65} for
$\tau_{\rm RE}$ observed in RE doped YIG as can be seen in the
inset of Fig.\ref{temp}. The shorter $\tau^{300 \ \rm K }_{\rm
RE}$ causes the field shift to be rather small at RT,
cf.~Fig.\ref{dh-ho}. As $\omega \tau_{\rm RE}\ll 1$ for all our
measurements Eq.~\ref{emori} predicts a linear frequency
dependence of the FMR linewidth in agreement with the experimental
results shown in Fig.~\ref{dh}.

In addition to the temperature dependence also the magnitude of
the RE induced linewidth is similar for YIG and Ni$_{80}$Fe$_{20}$
if one considers the ratio of RE concentration to the
magnetization. By scaling the YIG results from Dillon
\cite{Dillon-PR62} one expects  for a doping level of 2\% at room
temperature a linewidth of a few hundred Oe at 10~GHz for RE=Tb
and Dy in agreement with the present results; see Fig.~\ref{temp}.
It is therefore compelling to conclude that the additional damping
due to RE doping in Ni$_{80}$Fe$_{20}$ is caused by the very same
slow relaxing impurity mechanism which was originally proposed for
RE doped YIG.

Our experimental results sharply contradict earlier experimental
\cite{Reidy-APL03} and theoretical work \cite{Rebei-PRL03}. Reidy
{\it et al.}~\cite{Reidy-APL03} performed dynamic measurements
using only pulsed inductive magnetometry at very low frequencies
($\sim 500$~MHz). In addition these measurements suffer from a
large uncertainty with respect to the RE content of the samples.
The corresponding data points are included in Fig.~\ref{irkin}.
The absolute values for the damping constants derived from those
measurements are up to a factor of 5 lower than our present
results. However, we would like to point out that our present
results are in excellent agreement with earlier measurements for
Tb doped Ni$_{80}$Fe$_{20}$ films deposited under similar
conditions as the films used in the present study
\cite{Bailey-IEEE01,Russek-JAP02} (see Fig.~\ref{irkin}). Bailey
{\it et al.}~found a strong dependence of the contributed damping
on the Ar-pressure during the film deposition
\cite{Bailey-IEEE01}, with larger damping observed in films
deposited at lower sputter gas pressures. Lower sputter gas
pressures and lower deposition rates typically lead to smoother,
more homogeneous films with larger grain sizes. Note, that the
films used in the present study were deposited at even lower
sputter gas pressures and deposition rates than the ones in
\cite{Bailey-IEEE01}.  The theoretical work by Rebei and Hohlfeld
\cite{Rebei-PRL03} is based on orbit-orbit coupling effects
between RE impurities and itinerant electrons but does not
consider the slow relaxing impurity model. Rebei {\it et
al.}~solely justify the usage of itinerant electrons on the basis
of better agreement with the experimental data by Reidy {\it et
al.}~\cite{Reidy-APL03}. However, their theory results in
temperature independent Gilbert damping without a negative field
shift and strictly proportional to the orbital moment of the RE
impurities. These predictions are clearly at variance with our
experimental results. On the other hand the present results
(temperature, frequency, and element dependence of the RE induced
damping) can be readily explained by the slow relaxing impurity
model.

We would like to acknowledge financial support from the DFG through Sonderforschungsbereich
689 and priority program 1133. G.M.~gratefully acknowledges support from the
Alexander-von-Humboldt foundation.

\end{document}